\documentstyle[preprint,prabib,aps,12pt]{revtex}
\begin{document}
\title{Multidimensional SU(2) wormhole between two
null surfaces}
\author{ Dzhunushaliev V.D.
\thanks{E-mail address: dzhun@freenet.bishkek.su}}
\address {Department of the Theoretical Physics \\
          Kyrgyz State National University, Bishkek, 720024}
\maketitle
\begin{abstract}
{The  multidimensional  gravity  on  principal   bundle   with
structural $SU(2)$  gauge  group  is  considered.  The   Lorentzian
wormhole solution disposed between two null surfaces is  founded
in this case. As the nondiagonal components are similar to a gauge 
fields (in Kaluza - Klein's sense) then in some sense this solution 
is dual to the black hole in 4D Einstein - Yang - MIlls gravity: 
4D black hole has the stationary area outside of event horizon but 
multidimensional wormhole inside null surface (on which  $ds^2=0$)}
\end{abstract}

\section{Introduction}
\par
In gravity various physical phenomena can be  connected  with
wormholes ($WH$). In  quantum  gravity  they  form  {\it spacetime  foam}.
According to J. Wheeler, the classical $WH$  with  electrical  field
can describe the ``charge without charge''.  The  force  line  flows
into one side of $WH$ (``minus'' charge) and  outflow  on  other  side
(``plus'' charge). The $WH$ have been  studied  in  many  papers.  The
Euclidean $WH$ interacting with classical fields are  considered  in
\cite{gid}-\cite{cav}. In some works the multidimensional $WH$ 
are  examined  \cite{cl}-\cite{dzh1}.
In this article I shall receive the $WH$ in multidimensional gravity
in which the $SU(2)$ gauge group is  supplementary  coordinates.  In
such multidimensional gravity the total space is the fiber  bundle
with fiber~=~gauge group and base~$=~4D$ Einstein's spacetime. Since
any group is  symmetric  space,  its  metric  is  the  conformally
Euclidean metric and it can be written in the following view:
\par
\begin{equation}
ds^{2}_{fiber} = h(x^{\mu }) \sigma ^{a}\sigma _{a},
\label{1}
\end{equation}
where conformal factor $h(x^{\mu })$ depends only on spacetime coordinates
$x^{\mu }; \mu =0,1,2,3; \sigma _{a}=\gamma _{ab}\sigma ^{b}; \gamma _{ab}$  
is  Euclidean  metric; $a=4,5,\ldots N$
indexes of supplementary coordinates; $\sigma ^{a}$  are  one-form  
satisfies Maurer - Cartan structure equations:
\begin{equation}
d\sigma ^{a} = f^{a}_{bc}\sigma ^{b}\wedge\sigma ^{c},
\label{2}
\end{equation}
where $f^{a}_{bc}$ is a structural constant.

\section{$U(1)$ wormhole}

We remind the solution for $5D$ Kaluza - Klein's theory derived
in \cite{dzh1}. The metric is:
\begin{eqnarray}
ds^{2} = e^{2\nu (r)}dt^{2} - e^{2\psi (r)}(d\chi  - 
\nonumber\\
\omega (r)dt)^2 - dr^{2}-
a^2(r)(d\theta ^{2} + \sin ^{2}\theta  d\varphi ^2),
\label{3}
\end{eqnarray}
 where $\chi $ is the 5th supplementary coordinate; $r,\theta ,\varphi $  
are $3D$  polar
coordinates; $t$ is the time. The solution $5D$  Eistein's  equations
is:
\begin{eqnarray}
a^{2} & = & r^{2}_{0} + r^{2},
\label{4a}\\
e^{-2\psi } = e^{2\nu } & = & {2r_{0}\over q}{r^{2}_{0}+r^{2}
\over r^{2}_{0}-r^{2}},
\label{4b}\\
\omega & = & 4r^{2}_{0}\over q}
{r\over {r^{2}_{0} - r^{2}}.
\label{4c}
\end{eqnarray}
 This solution is the $WH$ between two null surfaces.

\section{$SU(2)$ wormhole}

We can introduce the Euler angles on $SU(2)$  group.  Then
one-forms $\sigma ^{a}$ can be written as a follows:
\begin{eqnarray}
\sigma ^{1} & = & {1\over 2}
(\sin \alpha d\beta - \sin \beta \cos \alpha d\gamma ),
\label{5a}\\
\sigma ^{2} & = & -{1\over 2}(\cos \alpha d\beta +
\sin \beta \sin \alpha d\gamma ),
\label{5b}\\
\sigma ^{3} & = & {1\over 2}(d\alpha +\cos \beta d\gamma ),
\label{5c}
\end{eqnarray}
where $0\le \beta \le \pi , 0\le \gamma \le 2\pi , 0\le \alpha \le 4\pi $. 
Thus, $7D$ metric on  the  total  space
can be written in the following view:
\begin{equation}
ds^{2} = ds^{2}_{fibre} + 2 G_{A\mu } dx^{A} dx^{\mu },
\label{6}
\end{equation}
 where $A=0,1,\ldots ,N$  is  multidimensional  index.  There  are   the
following  independent  degrees  of  freedom  in  multidimensional
gravity on the bundle: conformal factor $h(x^{\mu })$  and  components  of
the multidimensional metric $G_{A\mu }$.  Variation  of  the  action  with
respect to $G_{A\mu }$ leads to corresponding Einstein's equations:
\begin{equation}
R_{A\mu } -{1\over 2}G_{A\mu }R = 0,
\label{7}
\end{equation}
 where $R_{A\mu }$ is Ricci tensor and $R$ is Ricci scalar. Variation 
of the actions with respect to conformal factor $h(x^{\mu })$  is  
represented  in the following view:
\begin{equation}
\delta S = \delta G^{ab}{\delta S\over \delta G^{ab}}
\propto \delta h \gamma ^{ab}{\delta S\over \delta \gamma ^{ab}},
\label{8}
\end{equation}
 where $G_{ab}=h\gamma _{ab}$ is metric  on  fibre  of  the  bundle.  
Hence,  the corresponding equation has the following form:
\begin{equation}
R^{a}_{a} = R^{4}_{4} + R^{5}_{5} + R^{6}_{6} = 0.
\label{9}
\end{equation}
 Thus, the equations system described the  multidimensional  gravity
on principal bundle with fibre = nonabelian gauge group  are  (\ref{7}),
(\ref{9}) equations. If the gauge group is Abelian $U(1)$ group,  we  have
the standard $5D$ Kaluza - Klein's theory.
\par
We remind the following result \cite{per}-\cite{sal}. Let $G$  group  be  the
fibre of principal  bundle.  Then  there  is  the  one-to-one
correspondence between $G$-invariant metrics on the  total  space ${\cal X}$
and the triples $(g_{\mu \nu }, A^{a}_{\mu }, h\gamma _{ab})$. 
Where $g_{\mu \nu }$ is Einstein's pseudo  -
Riemannian metric; $A^{a}_{\mu }$ is gauge field  of  the $G$  group; 
$h\gamma _{ab}$  is symmetric metric on the fibre.
\par
We see a solution of the form:
\begin{eqnarray}
ds^{2} & = & e^{2\nu (r)}dt^{2}  - r^{2}_{0}e^{2\psi (r)}\sum^{3}_{a=1}
\left (\sigma ^{a} - A^{a}_{\mu }(r)dx^{\mu }\right )^{2} -
\nonumber\\
dr^{2} & - & a^{2}(r)\left (d\theta ^{2} + 
\sin ^{2}\theta d\varphi ^2\right ) .
\label{10}
\end{eqnarray}
We choose the ``potentials'' $A^{a}_{\mu }$ in following 
monopole-like form:
\begin{eqnarray}
A^{a}_{\theta } & = & {1\over 2}(f(r)+1)\{ \sin \varphi ;-\cos \varphi ;
0\} ,
\label{11a}\\
A^{a}_{\varphi } & = & {1\over 2}(f(r)+1)
\sin \theta \{\cos \varphi \cos \theta ;
\nonumber \\
& & \sin \varphi \cos \theta ;-\sin \theta \},
\label{11b}\\
A^{a}_{t} & = & v(r)\{ \sin \theta \cos \varphi ;
\sin \theta \sin \varphi ;\cos \theta \},
\label{11c}
\end{eqnarray}
Let  us  introduce  tetrads $e^{\bar{A}}_{A}$:
\begin{eqnarray}
ds^{2} & = & \eta _{\bar{A}\bar{B}}\Sigma ^{\bar{A}}
\Sigma ^{\bar{B}},
\label{12a}\\
\Sigma ^{\bar{A}} & = & e^{\bar{A}}_{A} dx^{A},
\label{12b}
\end{eqnarray}
where $\bar{A},\bar{B}=0,1,\ldots ,6$ are  tetrads  indexes; 
$\eta _{\bar{A}\bar{B}}$  is $7D$  Minkowski
metric. The input equations are written  below  in  the  following
form:
\begin{eqnarray}
R_{\bar{A}\mu } & = & 0,
\label{13a}\\
R^{a}_{a} & = & 0.
\label{13b}
\end{eqnarray}
$7D$ gravity equations become:
\begin{eqnarray}
& & \nu ''  + {\nu '}^2 + 3\nu ' \psi '  
+ 2{a' \nu ' \over a} -
{r^{2}_{0} \over 2} e^{2(\psi -\nu )}{v'}^2 -
\nonumber\\
& & {r^{2}_{0}\over a^{2}} v^{2}f^{2}e^{2(\psi -\nu )}  =  0,
\label{14a}\\
& & \nu ''  + {\nu '}^2 + 3\psi ''  + 
3{\psi '}^2 + 2{a'' \over a} -
{r^{2}_{0}\over 2} e^{2(\psi -\nu )}{v'}^2 +
\nonumber\\
& & {r^{2}_{0}\over 4a^2} {f'}^2e^{2\psi }  =  0,
\label{14b}\\
& & {a'' \over a} + {a'\over a}(\nu ' + 3\psi ' ) + 
{{a'}^2\over a^2}  - {1\over a^2}  + 
\nonumber\\
& & {r^{2}_{0}\over 8a^2} e^{2\psi }{f'}^2 -
{r^{2}_{0}\over 2a^2} v^2 f^2 e^{2(\psi -\nu )} +
\nonumber\\
& & {r^{2}_{0}\over 8a^{4}}\left (f^2 -1\right )^2  =  0,
\label{14c}\\
& & \psi ''  + 3{\psi '}^2 + 
2{a' \psi '\over a} + \psi ' \nu '  +
\nonumber\\
& & {r^{2}_{0}\over 6} e^{2(\psi -\nu )} {v'}^2 -
{2\over r^{2}_{0}} e^{-2\psi } - 
{r^{2}_{0}{f'}^2\over 12a^2} e^{2\psi } + 
\nonumber\\
& & {r^{2}_{0}\over 3}{v^2 f^2\over a^2}e^{2(\psi -\nu )} -
{r^{2}_{0}\over 24a^4} \left (f^2 -1\right )^2  =  0,
\label{14d}\\
& & f''  + f' (\nu ' + 5\psi ') + 
2v^2 f e^{-2\nu }  = 
\nonumber\\
& & {f\over 2a^{2}}(f^2 -1),
\label{14e}\\
& & v''  - v'(\nu ' - 5\psi ' - 2{a'\over a})  =  2{v\over a}f^{2},
\label{14f}
\end{eqnarray}
here the Eq's(\ref{14e})  and (\ref{14f})  are  ``Yang  -  Mills''  
equations  for
nondiagonal  components  of  the  multidimensional   metric.   For
simplicity we consider $f=0$ case. This means that  we  have  ``color
electrical'' field $A^{a}_{i}$ only (i=1,2,3). In this case it  is  easy  to
integrate Eq.(\ref{14f}):
\begin{equation}
v'  ={q\over r_{0}a^{2}} e^{\nu -5\psi },
\label{15}
\end{equation}
 where $q$ is the constant of  the  integration  (``color  electrical''
charge). Let us examine the most interesting case when the  linear
dimensions  of  fibers $r_{0}$  are  vastly  smaller  than  the  space
dimension $a_{0}$ and ``charge'' $q$ is sufficiently large:
\begin{equation}
\left ({q\over a_{0}}\right )^{1/2} \gg 
\left ({a_0\over r_{0}}\right )^2 \gg  1,
\label{16}
\end{equation}
 where $a_{0}=a(r=0)$ is the throat of the WH.
\par
On this approximation we deduce the equations system:
\begin{eqnarray}
& & \nu ''  + {\nu '}^2 + 
3\nu ' \psi '  + 2{a' \nu '\over a}  - {q^2\over 2a^4} e^{-8\psi } 
\nonumber\\
& &  =  0,
\label{17a}\\
& & \nu ''  + {\nu '}^2 + 3\psi ''  + 3{\psi '}^2 + 
2{a'' \over a} - {q^2\over 2a^4} e^{-8\psi }
\nonumber\\
& &   =  0,
\label{17b}\\
& & \psi ''  + 3{\psi '}^2 + 2{a' \psi '\over a} + 
\psi ' \nu '  + {q^2\over 6a^4} e^{-8\psi } 
\nonumber\\
& &  =  0,
\label{17c}\\
& & {a'' \over a} + {a'\over a}(\nu ' + 3\psi ' ) + 
{{a'}^2\over a^2}  - {1\over a^2} 
\nonumber\\
& &  =  0.
\label{17d}
\end{eqnarray}
 This system has the following solution:
\begin{eqnarray}
\nu & = & -3\psi,
\label{18a}\\
a^2 & = & a^2_0 + r^2,
\label{18b}\\
e^{-{4\over 3}\nu } & = & {q\over 2a_0}
\cos \left (\sqrt {8\over 3}\arctan {r\over a_0} 
     \right ),
\label{18c}\\
v & = & \sqrt 6{a_{0}\over r_0 q}
\tan \left (\sqrt {8\over 3} \arctan {r\over a_0}
     \right ).
\label{18d}
\end{eqnarray}
Let us define value $r$ in which  metric  has  null surfaces.  From
condition:
\begin{equation}
G_{tt}(r_g) = e^{2\nu (r_g)} - 
r^2_0 e^{2\psi (r)}\sum^3_{a=1}\left (A^a_t(r_g) \right )^{2} = 0
\label{19}
\end{equation}
it follows that:
\begin{equation}
{r_g\over a_0} = 
\tan \left (\sqrt {3\over 8}\arcsin \sqrt {2\over 3} 
     \right ) \approx  0.662.
\label{20}
\end{equation}
 It   is   easy   to   verify   that $\exp (2\nu )=0$ 
$(\exp (2\psi =\infty )$    by
$r/a_{0}=\tan (\pi \sqrt{3/32})\approx 1.434$. 
This  value  lies  beyond  the  null 
surfaces. This means that the small terms in (\ref{14c}), 
(\ref{14d}) will stay also small even near the null surfaces.

\section{Discussion}
One can concede that the  solution  of  this  sort  has certain
meaning in nature: it can be a source of the gauge fields.  
As  it  is  pointed  out  above,  there  is
one-to-one correspondence between the multidimensional  metric  on
bundle and $4D$ gravity + gauge field + scalar field. It is possible
that the mechanism of supplementary  coordinates  compactification
exists in nature. In  Ref.\cite{dzh2}  the  possible  mechanism  of  this
phenomenon is discussed from the algorithmical point of  view.  In
this case the composite $WH$ realizing J.Wheeler's idea  on  ``charge
without charge'' and ``mass without mass'' can exists in nature. Such
$WH$ can be constructe in the following  way:  The  multidimensional
region is found in center  of  the $WH$  and  two $4D$  regions  are
disposed on each side of center.  It  can  be  supposed  that  the
compactification happens on some chosen surface. The event horizon
can be such surface. Such composite $WH$  in $5D$  Kaluza  -  Klein's
theory was examined in \cite{dzh1}. Thus,  the  exterior  observer  cannot
detects the presence of this multidimensional  region  under  event
horizon, he can registers only electrical charge, angular  momentum
and mass of the black hole. It should be noted that in this  model
the compactification of the supplementary coordinates is a quantum
phenomenon (this is a jump and not a classical step -  by  -  step
splitting off the supplementary coordinates).


\begin{thebibliography}{20}
\bibitem{gid}
S.B. Giddings and A.Strominger, Nucl.Phys. ${\sl B306}, 890(1988)$.
\bibitem{my}
R.C.Myers, Phys.Rev.{\sl D38}, 1327(1988).
\bibitem{haw}
S.W.Hawking, Phys.Lett.{\sl B195}, 337(1987).
\bibitem{hos}
A.Hosoya and W.Ogura, Phys.Lett.{\sl B226}, 117(1989).
\bibitem{cav}
M.Cavagli\'a,  V.de  Alfaro  and  F.  de   Felice,   Phys.Rev.{\sl D49},
6493(1994).
\bibitem{cl}
A.Chodos, S.Detweiler, GRG, ${\sl 14}$, 879(1982);
G.Cl$\acute e$ment, GRG, ${\sl 16}, 477(1984)$.
\bibitem{dzh1}
V.D.Dzhunushaliev, Izv.  Vuzov,  ser.Fizika, $N6$, 78(1993).
\bibitem{per}
R.Percacci, J.Math.Phys., {\sl 24}, 807(1983).
\bibitem{sal}
A.Salam and J.Strathdee, Ann.Phys., {\sl 141}, 316(1982).
\bibitem{dzh2}
V.D.Dzhunushaliev, Izv. Vuzov,  ser.Fizika, $N9$, 55(1994).
\end{thebibliography}
\end{document}